\documentstyle[12pt, preprint]{aastex}

\def\ninth{\textstyle {1 \over 9} \displaystyle}
\def\fiveninth{\textstyle {5 \over 9} \displaystyle}
\def\ltsima{$\;\buildrel < \over \sim \;$}
\def\simlt{\lower.5ex \hbox{\ltsima}}
\def\gtsima{$\;\buildrel > \over \sim \;$}
\def\simgt{\lower.5ex \hbox{\gtsima}}
\received{2002 May 22}
\begin{document}
\title{SWAS and Arecibo observations of 
H$_2$O and OH in a diffuse cloud along the line-of-sight to W51}

\author{David A. Neufeld\altaffilmark{1}, Michael J. Kaufman\altaffilmark{2}, 
Paul F. Goldsmith\altaffilmark{3}, David J. Hollenbach\altaffilmark{4}
and Ren\'e Plume\altaffilmark{5}}

%\vskip 0.5 true in \parskip 6pt
  
\altaffiltext{1}{Department of Physics \& Astronomy,  The Johns Hopkins University, 3400
North Charles Street,  Baltimore, MD 21218}

\altaffiltext{2}{Department of Physics, San Jose State University,
One Washington Square, San Jose, CA 95192}

\altaffiltext{3}{NAIC, Department of Astronomy, Cornell
University, Ithaca, NY 14853}

\altaffiltext{4}{NASA Ames Research Center, MS 245-3, Moffett Field, CA 94035}

\altaffiltext{5}{Department of Physics \& Astronomy,
University of Calgary, 2500 University Drive N.W., Calgary, Canada}

%\vskip 0.1 true in

\keywords{ISM: abundances -- ISM: molecules -- ISM: clouds -- molecular processes -- 
radio lines: ISM -- submillimeter}

\begin{abstract}

Observations of W51 with {\it Submillimeter Wave
Astronomy Satellite} (SWAS) have yielded
the first detection of water vapor in
a diffuse molecular cloud.  The water vapor
lies in a foreground cloud that gives rise 
to an absorption feature at an LSR velocity 
of 6 km s$^{-1}$.  The inferred water column
density is $2.5 \times 10^{13} \, \rm cm^{-2}$.
Observations with the Arecibo radio telescope
of hydroxyl molecules at ten positions in W51
imply an OH column density of 
$8 \times 10^{13} \, \rm cm^{-2}$ in the
same diffuse cloud.  The observed H$_2$O/OH
ratio of $\sim 0.3$ is significantly larger
than an upper limit derived previously from
ultraviolet observations of the similar 
diffuse molecular cloud lying in front of
HD 154368.  The observed variation in
H$_2$O/OH likely points to the presence
in one or both of these clouds
of a warm ($T \simgt 400$) gas component
in which neutral-neutral reactions are
important sources of OH and/or H$_2$O.

\end{abstract}

\section{Introduction}

Since its launch in December 1998, the {\it Submillimeter
Wave Astronomy Satellite} (SWAS; Melnick et al.\ 2000) 
has detected water vapor in
more than 70 molecular clouds by means of observations 
the $1_{10} - 1_{01}$ transition of ortho-H$_2$O (e.g.
Snell et al.\ 2000; Ashby et al.\ 2000; Neufeld et al.\ 2000a). 
While emission-line observations form the core of the SWAS program 
on interstellar water vapor,  absorption-line observations are 
possible toward a few bright continuum sources; these include 
Sagittarius B2 (Neufeld et al.\ 2000b; hereafter N00), 
Sagittarius A, W49, and W51.  Absorption-line observations 
typically probe the water vapor abundance in several 
kinematically-distinct foreground clouds lying along the 
line-of-sight to the source (e.g. N00). 
Under typical interstellar conditions, most water molecules are
in the lower state of the $1_{10} - 1_{01}$
transition (i.e. in the ground state of ortho-water); thus 
absorption line observations have the distinct advantage of 
yielding water vapor column densities that are insensitive 
both to the physical conditions in the absorbing
cloud and to the assumed rate coefficients for collisional
excitation of water.  Typically, the H$_2^{16}$O absorption 
line is very optically-thick, yielding 
only a lower limit on the water column density, but 
observations of {\it optically-thin} absorption by the 
H$_2^{18}$O isotopologue (less abundant by a factor 
of 250--500) have led to a quantitative determination
of the water column densities in foreground clouds along
the Sgr B2 sight-line (N00).

In this {\it Letter}, we report the results of similar
observations carried out towards the star-forming
region W51.  The results are  particularly intriguing, because 
they provide our first detection of a H$_2^{16}$O 
absorption line which is of only {\it moderate} optical depth.
This feature, observed at an LSR velocity $\sim 6$~km~s$^{-1}$,
originates in a diffuse foreground cloud in the which the
water column density is small.  This cloud has 
previously been detected by means of 21~cm absorption-line
observations (Koo 1997): the inferred HI column density is $\sim
10^{21}\,\rm cm^{-2}$.

Spaans et al. (1998; hereafter
S98) have argued that measurements of OH and H$_2$O column
densities in diffuse clouds provide a valuable probe of
the chemistry of interstellar oxygen molecules; in particular, 
the OH/H$_2$O abundance ratio serves to constrain the
branching ratio for the dissociative recombination of 
the molecular ion H$_3$O$^+$, a crucial parameter in 
chemical models for both diffuse and dense molecular clouds.
Accordingly, we have used the Arecibo 
Observatory (AO) to carry out OH absorption line 
observations toward the same source, the results of 
which are also presented here.

The observations and data reduction are described in \S 2 below, 
and the observational results presented in \S 3.  In \S 4 we 
discuss the derived water and OH column densities and the 
constraints that they place upon the oxygen chemistry in 
molecular clouds.

\section{Observations}

Our SWAS observations of W51 were carried out during the period 
1999 April 16 -- 2001 April 19 with the 
$3.3^\prime \times 4.5^\prime$ elliptical SWAS beam centered
at position $\rm \alpha = 19^h\, 23^m \, 43.^s0$,
$\delta= 14^\circ 30^\prime 38^{\prime\prime}$ (J2000).  
All the data were acquired
in standard nodded observations (Melnick et al.\ 2000) and 
were reduced using the standard SWAS pipeline.  The total 
on-source integration time was 58.6 hours.

AO observations of the 1612, 1665, 1667, and 1720 MHz lines 
of OH were obtained on 2001 November 30 and 2001 December 3
towards the SWAS-observed position and nine offset positions
of widely varying continuum flux.  The goal of observing OH
towards these offset positions was to
allow inferences to be drawn about the line excitation 
temperatures (see \S 3.2 below).
The offsets and on-source integration times for each observed
position are given in Table 1, all positional offsets
being expressed relative to the SWAS-observed position.
The AO data were obtained
in a series of standard nodded observations in which the
source and reference positions were alternately observed for 
a period of 5 minutes each.  The reference position was
chosen to have an R.A. offset of --5~(cos$\,\delta$)$\,$ minutes of time
so that the source
and reference positions were observed over identical tracks
in azimuth and elevation.  Calibration measurements
were carried out using the noise diode after each source/reference 
position pair, and calibration observations of an astronomical
continuum source were performed at the beginning of each
day's observations.

\section{Results}

\subsection{SWAS observations of water vapor}

Figure 1 shows the complete spectrum of the $1_{10}-1_{01}$
556.936 GHz transition of H$_2$O obtained by SWAS toward W51.  
The continuum antenna temperature, $T_{AC}^*$,
measured toward W51 was 0.35~K (double sideband),
corresponding to a 550 GHz continuum flux density of $2.5 \times 10^3$~Jy.
The quantity plotted in Figure 1 is the ratio of flux density to 
continuum flux density, given by
$[T_A^* - 0.5\, T_{AC}^*] / \, 0.5 \, T_{AC}^*$
under the assumption that the sideband gains are equal.
The SWAS beam size is $3.3^{\prime} \times 4.5^{\prime}$ (FWHM) at 550 GHz.
The fitted line at $v_{\rm LSR} \sim 6$~km~s$^{-1}$
is the best-fit absorption line for a
cloud that is assumed to completely cover the source. 
The derived cloud parameters are a line center optical
depth $\tau_0$ of 1.5, a Doppler parameter\footnote{
The Doppler parameter $b$ is defined (as usual) 
such that the optical depth drops by a factor
e=2.718...  at a velocity shift $b$ from line center.
The corresponding FWHM obtained from a Gaussian fit to 
the absorption line is
3.4~km~s$^{-1}$.} $b$ of 1.5~km~s$^{-1}$,
and cloud LSR velocity of 6.0~km~s$^{-1}$.  Under
the conditions typical of diffuse interstellar clouds,
the population in the $1_{10}$ rotational state
of water is very small and thus the effects of
stimulated emission on the $1_{10}-1_{01}$ optical
depth can be neglected.   The column
density of water in the 1$_{01}$ state is straightforwardly
obtained as $1.9 \times 10^{13}$~cm$^{-2}$.
If the ortho-para
ratio for water is 3, the implied total water column density
is $2.5 \times 10^{13}$~cm$^{-2}$.  The water abundance
relative to HI is $\sim 10^{-8}$.

\subsection{Arecibo observations of OH}

Figure 2 shows the OH spectra obtained at AO towards
the ten positions observed in W51.  
These spectra show the ratio of beam-averaged brightness 
temperature, $T_B$, to continuum
brightness temperature, $T_{BC}$.  The beam-averaged
brightness temperature of the radiation incident upon 
the AO antenna is
computed according to $T_{B} = T_A / \eta_A + T_{\rm CMB}$,
where $T_A$ is the antenna temperature, $\eta_A \sim  0.8$ 
is the aperture efficiency -- including atmospheric
losses, as determined from calibration observations of an 
astronomical continuum source -- 
and $T_{\rm CMB} = 2.73$~K is the temperature of the cosmic background 
radiation (which is ``chopped out" by our observing 
procedure).  The size of the Arecibo main beam is 
$\sim 2.6^{\prime} \times 3.0^{\prime}$ (FWHM) at 1666 MHz.

The continuum brightness temperatures, $T_{BC}$,
measured for each position are shown in Table 1, along
with the measured equivalent width, $W_v$ -- in units of km s$^{-1}$
-- for each of the observed OH lines.  Here,
$W_v$ is defined as $\int (T_B - T_{BC}) / T_{BC}] \, dv$, 
so that a negative value
of $W_v$ implies an absorption line and a positive value an
emission line. 
Because of strong
radio frequency interference, the 1720 MHz line strength could
be measured reliably only toward the SWAS-observed (0,0) position
where the absolute line strength was largest.

The results shown in Table 1 
imply that the ratio of equivalent widths
departs greatly from the 1:5:9:1 value expected
in local thermodynamic equilibrium (LTE) for
$W_v(1612): W_v(1665): W_v(1667): W_v(1720)$.  
In particular,
the 1665 MHz / 1667 MHz absorption line ratio
is considerably smaller than the LTE value and
the 1612 MHz / 1667 MHz ratio considerably larger.
Moreover, the 1720 MHz line is observed in emission
toward the (0,0) position, indicating the presence
of weak maser amplification.  
  
For lines of small optical depth, 
the equivalent width is
related to the OH column density and the
excitation temperature, $T_{ex}$, by the expression
\begin{equation}
W_v = 0.45 \, k\, [N({\rm OH})/ 10^{14}{\rm cm^{-2}}] \, [T_{ex}/{\rm K}]^{-1} \,
[1 - T_{ex}/ T_{BC}] \,
\rm km \, s^{-1} \\
\end{equation}
where $k$ = $\ninth$, $\fiveninth$, 1 and $\ninth$
respectively for the 1612, 1665, 1667 and 1720 MHz transitions.
The ratios and signs of the equivalent widths observed toward
the (0,0) position
imply that $T_{ex}(1720) < 0 < T_{ex}(1612) < T_{ex}(1667)
< T_{ex}(1665)$.  Assuming that all the excitation temperatures
are small compared to $T_{BC}$ at the (0,0) position (i.e.
small compared to 500~K),
we find the ratio of excitation temperatures
to be  $T_{ex}(1612): T_{ex}(1665) : T_{ex}(1667): T_{ex}(1720) 
= 0.18: 1.79: 1: -0.29$.  

Unfortunately, because the equivalent
widths are a function of {\it two} unknowns -- $N({\rm OH})$ 
and $T_{ex}$ -- the OH column density and excitation
cannot be derived independently for any given sight-line.
However, as we argue below, the variation of the equivalent
widths with the background continuum brightness temperature,
$T_{BC}$, for the ten sight-lines we observed,
provides a strong argument that the entire region is
covered by a large foreground cloud in which the OH
column density and excitation are nearly constant.  This
in turn allows us to derive an estimate of $N({\rm OH})$.

In Figure 3 (upper panel), we show the equivalent widths measured 
for each transition and each observed position as a function
of the background continuum brightness temperature,
$T_{BC}$.  We note immediately that the equivalent
width of the 1612 MHz transition shows rather little
variation over the entire set of observed positions;
$W_v(1612)$ lies within $30\%$ of 0.08~km~s$^{-1}$ for
every sight-line that we observed.  Referring to
equation (1), we see that the near-constancy of $W_v$
argues strongly that $N({\rm OH})$ and $T_{ex}$ are nearly
constant and that $T_{ex}$ is small compared to $6$~K,
the smallest $T_{BC}$ for any sight-line that we observed.
In principle, of course, $N({\rm OH})$ and $T_{ex}$ could
show large variations from one position to another, but
it would seem highly improbable that such variations
could conspire to yield a right-hand-side for equation (1)
that is nearly constant.  Therefore, we shall henceforth
assume that the OH column density and excitation are nearly the same
towards each of the ten positions we have observed.

It is also apparent from Figure 3 that the equivalent
widths of the 1665 MHz and perhaps the 1667 MHz transition
appear to show a systematic decline in those sight-lines 
of smallest continuum brightness; this decline allows
the line excitation temperatures to be estimated.
In Figure 3 (lower panel), we plot the ratio of equivalent widths
for the 1665 and 1612 MHz transitions, again as a function
of the background brightness temperature.  The solid curve
shows the best fit to the data, for a model in which the
1612 and 1665 excitation temperatures are assumed to be
constant and in the 0.1:1 ratio inferred for the (0,0)
position.  The best fit curve is obtained for 
$T_{ex}(1665) = 7~K$.  Adopting this value for $T_{ex}(1665)$,
we infer from equation (1) and the observed 1665 MHz equivalent
width that 
the OH column density is $8 \times 10^{13}$~cm$^{-2}$
towards the SWAS-observed (0,0) position.  

\section{Discussion}

Because of the relative simplicity of the chemical networks
involved, diffuse molecular 
clouds provide a useful laboratory for testing astrochemical
models; in particular, the H$_2$O/OH abundance ratio
serves as a valuable probe of the chemical network that
produces oxygen-bearing molecules (S98).
One key uncertainty in that network concerns the dissociative
recombination of H$_3$O$^+$ with electrons, and specifically the fraction
of such recombinations that produce water, $f_{\rm H_2O}$,
a quantity for which two laboratory groups have obtained 
highly discrepant results. 
According to results obtained in the flowing afterglow experiment
of Williams et al.\ (1996), 
a fraction $f_{\rm OH}=0.65$
of dissociative recombinations of H$_3$O$^+$ lead to
OH, a fraction $f_{\rm H_2O} = 0.05 $ to H$_2$O, and the
remaining fraction 
$f_{\rm O}= 1 - f_{\rm OH} - f_{\rm H_2O}$ to O.
A different experimental technique 
(Vejby-Christensen et al.\ 1997), which
made use of the ASTRID heavy-ion storage ring in Denmark, 
yielded significantly different results (Jensen et al. 2000), {\it viz.}
$f_{\rm OH} : f_{\rm H_2O}: f_{\rm O} = 0.74 \pm 0.02 : 0.25 \pm 0.01 : 0.013 \pm 0.005$.
Similar results (although with larger error bars) were obtained (Neau et al.\ 2000)
from the CRYRING heavy-ion storage ring facility; they were 
$f_{\rm OH} : f_{\rm H_2O}: f_{\rm O} = 0.78 \pm 0.08 : 0.18 \pm 0.05 : 0.04 \pm 0.06$,

Taken together with the 
ground-based observations of OH that we obtained
at Arecibo, the SWAS observations of W51
imply an H$_2$O/OH abundance ratio $\sim 0.3$ in the
diffuse cloud that is responsible for the
$v_{\rm LSR} = 6 \rm \, km \, s^{-1}$ feature. 
Considering the uncertainties in our
determination of the H$_2$O and (particularly) the
OH column densities, we
estimate the H$_2$O/OH ratio to be uncertain by a 
factor $\sim 2$.
In comparing 
the observed H$_2$O/OH abundance ratio with theoretical
predictions, we have used the steady-state photodissociation region
(PDR) model of Kaufman et al.\ (1999), modified so as to treat the case
of a finite slab illuminated from two sides.  Because H$_2$ and CO 
are photodissociated following line absorption, their photodissociation
rates are reduced by self-shielding.  In order to treat correctly
the effects of self-shielding for radiation incident
upon {\it both} sides of the slab, the H$_2$ and CO abundances must be 
obtained by an iterative method. 

\subsection{Standard models of cold diffuse clouds}

In Figure 4, we show the predicted H$_2$O and OH column
densities for a variety of {\it astrophysical} parameters: 
the total visual extinction through the cloud, $A_V$, in magnitudes;
the illuminating UV field, $G_0$, in units of the Habing field; 
and the cosmic ray ionization rate, $\zeta_{\rm cr}$.  
All results apply to an assumed cloud density, $n_H$, of 
100 H nucleons per cm$^{3}$.  The temperature is calculated
from considerations of thermal balance and is $\sim 30$~K at
the cloud center.
Filled squares apply to models 
with $f_{\rm OH} : f_{\rm H_2O}: f_{\rm O}=0.75 : 0.25 :  0.0$ (values
suggested by the ASTRID storage ring experiment), while 
filled triangles apply to models with 
$f_{\rm OH} : f_{\rm H_2O}: f_{\rm O}= 0.65 : 0.05 :  0.30$ 
(suggested by the flowing afterglow experiment).
The different astrophysical
parameters for each plotted data point are described
in the figure caption.  Black circles represent the column densities
observed toward W51 and HD 154368.  

A striking feature of Figure 4 is that although the H$_2$O
and OH column densities depend strongly on the assumed
astrophysical parameters, their {\it ratio} is determined
primarily by the assumed branching ratio $f_{\rm H_2O}$ and 
shows almost no dependence upon $A_V$, $G_0$, or
$\zeta_{\rm cr}$.  This behavior can be understood by means
of a simple ``toy" model, in which we assume OH and H$_2$O
to be formed by dissociative recombination 
of H$_3$O$^+$ and destroyed by photodissociation.  
The expected H$_2$O/OH ratio 
is given by \begin{equation}
{n({\rm H_2O}) \over n(\rm OH)} = {\zeta_{\rm OH} \over \zeta_{\rm H2O}}
\times {f_{\rm H_2O} \over f_{\rm OH} + f_{\rm H_2O}} \sim 
0.69 \times {f_{\rm H_2O} \over f_{\rm OH} + f_{\rm H_2O}}
\end{equation}
where $\zeta_{\rm OH} = 3.5 \times 10^{-10} G_0 \exp(-1.7\,A_V) \,\rm s^{-1}$
and $\zeta_{\rm H2O} = 5.1 \times 10^{-10} G_0 \exp(-1.8\,A_V) \,\rm s^{-1}$
are the assumed photodissociation rates for OH and H$_2$O (Roberge
et al.\ 1991).\footnote{The 
quantity $f_{\rm H_2O}$ appears with $f_{\rm OH}$ in the denominator
of the second term on the right-hand-side, because photodestruction of
H$_2$O results in the formation of OH.}  Equation (2) yields H$_2$O/OH
abundance ratios of 0.172 and 0.049 respectively for the branching
ratios assumed for the filled squares and triangles in Figure 4.
These ratios are shown by dashed lines in Figure 4, and do indeed 
yield good agreement with results from the full steady-state PDR model. 
The assumption of chemical steady-state is justified by the fact
that the photodissociation timescale for OH is only $\sim 100 
\exp(1.7\,A_V) / G_0$ years.

Given standard models for cold diffuse clouds,  
the observed H$_2$O/OH abundance ratio of $\sim 0.3$ in W51 is 
{\it consistent} with the case $f_{\rm OH} : f_{\rm H_2O}: f_{\rm O} = 0.75 : 0.25 : 0$
and clearly {\it inconsistent} with the case $f_{\rm OH} : f_{\rm H_2O}: f_{\rm O} 
= 0.65 : 0.05: 0.30$.  
Thus -- if interpreted using standard models for cold diffuse clouds --
the observed H$_2$O/OH abundance ratio
argues for the laboratory results obtained in the ASTRID storage ring
experiment and
against those obtained in the flowing afterglow experiment.

This conclusion, however, is different from that obtained 
by S98, who used ultraviolet
absorption line observations with the Goddard High Resolution
Spectrometer of HST to place an upper limit of only 0.06 ($3\,\sigma$) 
on the H$_2$O/OH ratio in an entirely different diffuse cloud,
which lies in front of the star HD 154368.  Based upon these observations, S98 
argued for a {\it low} value of $f_{\rm H_2O}$ that is inconsistent
with the ASTRID storage ring experiment.  The puzzling
discrepancy between the observed H$_2$O/OH abundance ratio
in these two diffuse clouds
may point to the importance of additional production 
mechanisms for OH or H$_2$O that do not involve
the dissociative recombination of H$_3$O$^+$.  This possibility is addressed
in \S4.2 below. 

\subsection{Enhanced-temperature models of diffuse molecular clouds}

It has long been recognized (e.g. Elitzur \& de Jong 1978) that 
neutral-neutral reactions provide an alternate production
route to OH and H$_2$O.  The reactions
\begin{equation}
\rm O + H_2  \rightarrow OH + H \\
\end{equation}
\begin{equation}
\rm OH + H_2  \rightarrow H_2O + H \\
\end{equation}
possess activation barriers that make them negligibly
slow at the low temperatures typical of diffuse
interstellar clouds;  at temperatures
above $\sim 300$~K, however, they become
important production mechanisms for OH and H$_2$O
(Neufeld, Lepp \& Melnick 1995).
If even a small fraction of the gas in the 
W51 $\rm 6\, km\, s^{-1}$ and/or the HD 154368 cloud
were sufficiently warm -- as a result of a weak shock,
for example -- then neutral-neutral reactions might
perturb the OH and H$_2$O column densities. 

To investigate this possibility, we have obtained model
predictions for PDRs in which the temperature has
been fixed at a variety of temperatures between 100 
and 1500~K.  The results are represented
by the magenta locus in Figure 4; they were obtained 
for the astrophysical 
parameters adopted by S98 for the HD 154368 cloud --   
$A_V = 2.65$~mag, $n_H = 325\,\rm cm^{-3}$, $G_0 = 3$ --
but for a branching ratio $f_{\rm H_2O}  = 0.25$ 
and a cosmic ray ionization rate of $1.8 \times 10^{-17}
\rm cm^{-3}$.  The OH and H$_2$O column densities are both clearly
enhanced by neutral-neutral reactions.  At moderate
temperatures, the H$_2$O/OH ratio decreases because
the large O/OH ratio makes reaction (3)
more important than (4).  At temperatures
higher than $\sim 600$~K, however, the effect on
the ratio is reversed, and the H$_2$O/OH ratio is 
{\it increased } by neutral-neutral reactions.

The results obtained in enhanced-temperature models
suggest a way out of the puzzle posed by the
discrepant OH/H$_2$O ratios measured in
the W51 $\rm 6\, km\, s^{-1}$ and HD 154368 
clouds.  If these clouds possess small (but differing)
amounts of warm gas, then the OH/H$_2$O abundance
ratios could differ (even though the value of $f_{\rm H_2O} $
must, of course, be identical in both clouds). 
We are unaware of any observations that rule out
the presence of small amounts of warm gas in these 
sources; indeed, the presence of such gas in diffuse clouds
is predicted by certain models that seek to explain
the anomalously-high CH$^+$ abundances observed in many
diffuse clouds as resulting from the effects of 
turbulence or weak shocks\footnote{Depending upon
the geometry, velocity shifts of the
OH and H$_2$O lines relative to the lines of other
species (e.g. HI) might be an observational signature of
a shock production mechanism.  The velocity shifts
can be very small, however, if the shock propagates
at an oblique angle to the line-of-sight or if
multiple shocks are present in the beam.  Thus, the
absence of any such signature in the W51 6 km $s^{-1}$
cloud does not argue strongly against the
production of OH and H$_2$O in shocks.} (e.g. 
Joulain et al.\ 1998, Flower \& Pineau des Forets\ 1998,
and references therein).  

Unfortunately -- as discussed above --
the effect of warm gas upon the H$_2$O/OH ratio
depends critically upon the gas temperature --
and even switches sign at $T\sim 600$~K.  Thus, if warm
gas is present, the observed H$_2$O/OH ratio
cannot even be used to derive a limit upon
$f_{\rm H_2O} $.  For example,
if $\sim 5\%$ of the gas in HD 154368 were at $\sim 500$~K,
then the observations of OH and H$_2$O in that source
could be reconciled with the large branching ratio 
$f_{\rm H_2O}  = 0.25$ derived in \S4.1 above.  Alternatively, 
if $\sim 0.3\%$ of the gas in the W51 cloud
were at $T\sim 700$~K, then the observed H$_2$O/OH ratio  
would be consistent with the lower limit on 
$f_{\rm H_2O} $ inferred previously by S98.

To summarize, the discrepancy between the H$_2$O/OH ratio
reported here for the W51 6 km s$^{-1}$ cloud 
and that reported previously for the HD 154368 cloud
(S98) suggests that a component of 
warm ($T \simgt 400$) gas is present in one or both 
of these sources.  The presence of
this warm component makes it difficult to determine
observationally the branching ratio for the
dissociative recombination of H$_3$O$^+$
with electrons to form OH and H$_2$O.
Our new observations of W51
cast doubt upon the conclusion of 
S98 that the branching ratio to H$_2$O is small,
but do not allow the branching ratio to be determined
definitively.   

This work was supported NASA's SWAS contract NAS5-30702.  
We gratefully acknowledge the excellent support of
the telescope operators at the Arecibo Observatory.
The Arecibo Observatory is operated by the National Astronomy
and Ionosphere Center under a Cooperative Agreement with
the National Science Foundation.

\clearpage

\begin{figure}
\epsscale{0.80}
\plotone{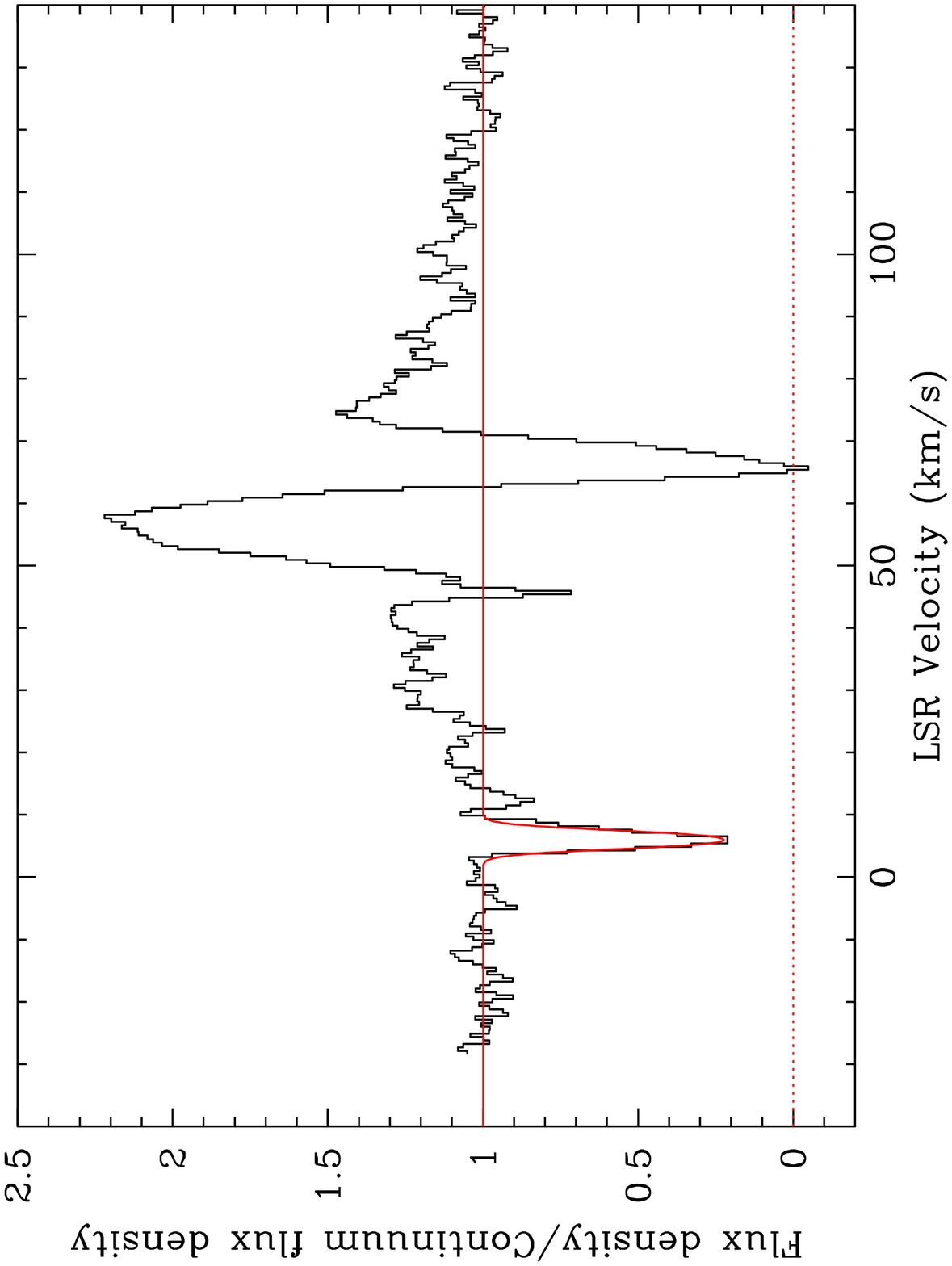}
\caption{Water vapor spectra obtained toward W51 by
the Submillimeter Wave Astronomy Satellite.  The figure shows the
$1_{10}-1_{01}$ pure rotational transition of $\rm H_2^{16}O$ near 
557~GHz.}
\end{figure}

\clearpage

\begin{figure}
\epsscale{0.80}
\plotone{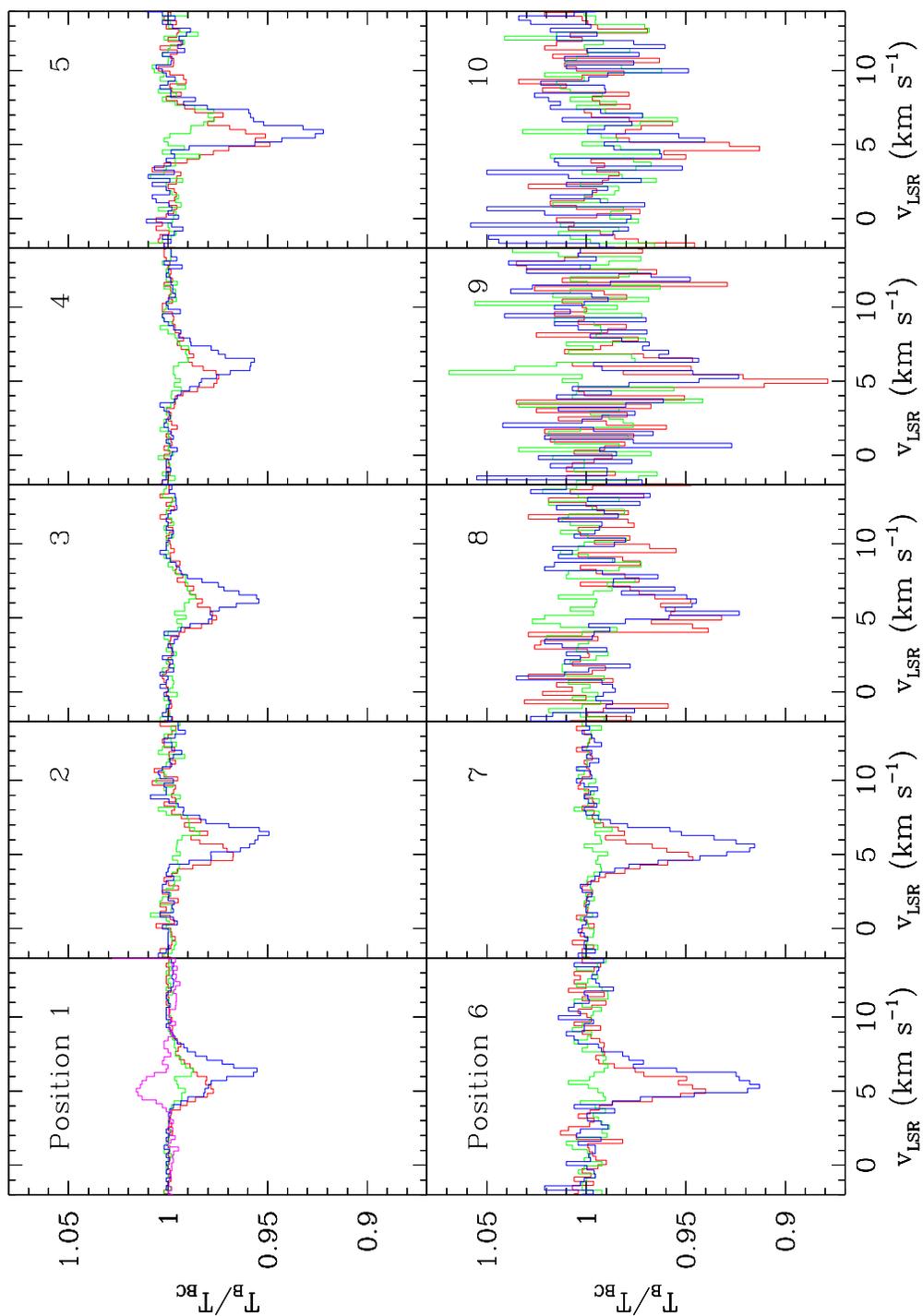}
\caption{OH spectra obtained toward ten positions in
W51.  The red, green, blue and magenta curves respectively show
the 1612, 1665, 1667, and 1720 MHz transitions.}
\end{figure}

\clearpage

\begin{figure}
\epsscale{0.80}
\plotone{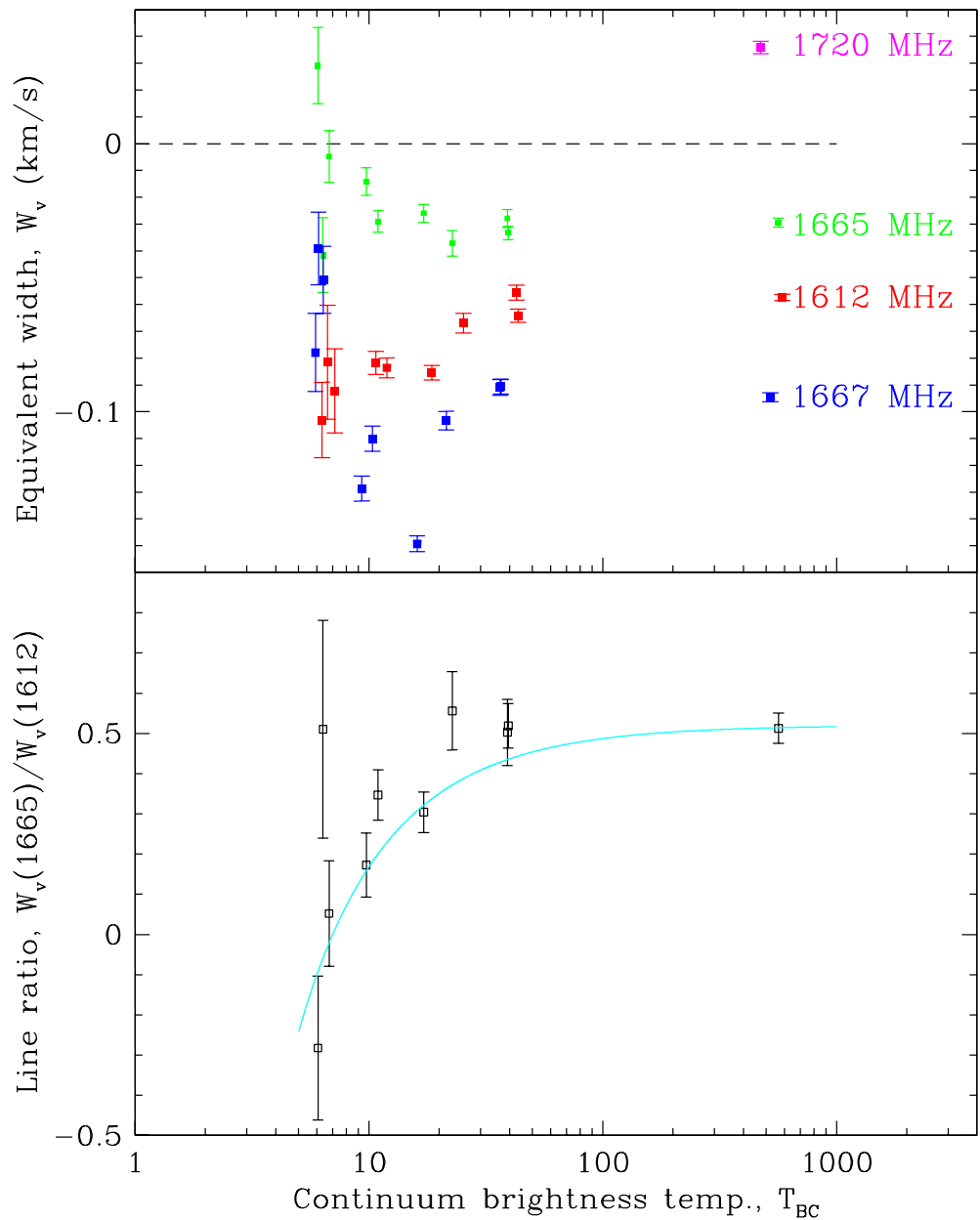}
\caption{OH equivalent widths observed toward ten positions
in W51 (upper panel), as a function of continuum brightness
temperature at each position.  
The lower panel shows the ratio of the 
equivalent widths for the 1665 MHz and 1612 MHz transitions.
The curve refers to a fitted model in which the foreground
OH column density and excitation is assumed to
be the same for all observed positions (see text for
the best-fit model parameters).}
\end{figure}
\clearpage

\begin{figure}
\epsscale{0.7}
\plotone{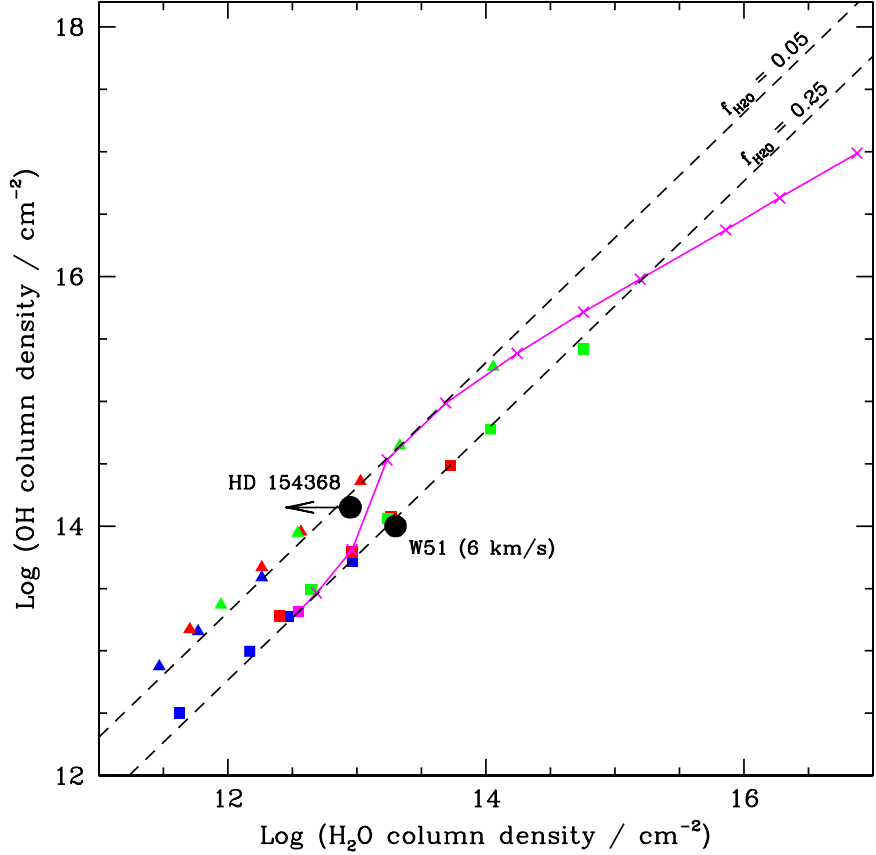}
\caption{OH and H$_2$O column densities predicted
by a photodissociation model.  Results have been
obtained for two different assumptions about the
branching ratio for dissociative recombination of
H$_3$O$^+$;  filled squares apply to models 
with $f_{\rm OH} : f_{\rm H_2O}: f_{\rm O}=0.75 : 0.25 :  0.0$ 
(results from the ASTRID storage ring experiment), while 
filled triangles apply to models with 
$f_{\rm OH} : f_{\rm H_2O}: f_{\rm O}=0.65 : 0.05 :  0.30$ 
(flowing afterglow experiment).
Red symbols shows the results of models in which the incident
UV field, $G_0$, is 5 and the cosmic ray ionization rate, $\zeta_{cr}$,
is $\rm 1.0 \times 10^{-16}\,s^{-1}$.  Blue and green 
symbols apply respectively to the cases $(G_0=5, \zeta_{cr}
=\rm 1.8 \times 10^{-17}\,s^{-1})$ and $(G_0=1, \zeta_{cr}
=\rm 1.0 \times 10^{-16}\,s^{-1})$.  In each case, results
were obtained for total cloud extinctions, $A_V$, of 1, 2,
3, and 4 (from left to right as the points appear in the
figure).  The black dashed lines show the H$_2$O/OH ratios
of 0.172 and 0.049 obtained for the two assumed branching 
ratios using the toy model described in the text.  The magenta
locus shows the results of enhanced
temperature models (see text): from left to right, the
points refer to a model in which the temperature
is calculated self-consistently (filled square; $T \sim 30$~K at 
cloud center) and then (crosses) 
to models in which the assumed temperature is fixed at 
100, 300, 400, 450, 500, 550, 600, 700, 800, and 1500~K.}
\end{figure}

\clearpage

%\vfill\eject 
{\scriptsize 
\begin{tabular}{c c c c c c c c} 
\multicolumn{8}{c}{TABLE 1}\\
\multicolumn{8}{c}{OH observations toward W51}\\
\\
\hline 

Position & Offset$^a$ & Time $^b$  & $T_{\rm BC}\,\,^c$ &  $W_v (1612)\,\,^d$ & $W_v (1665)$ & $W_v (1667)$ & $W_v (1720)$ \\
         &            &  (s)   & (K) &($\rm km\,s^{-1}$) 
& ($\rm km\,s^{-1}$) &($\rm km\,s^{-1}$) &($\rm km\,s^{-1}$) \\
\hline
\\

 1 & (0, 0)& 600 & 542.8 & --0.057 (0.001) & --0.029 (0.002) & --0.095 (0.002)& +0.036 (0.002) \\
 2 & (--1.3, --8.3)  & 600 &  22.1 & --0.067 (0.004) & --0.037 (0.005) & --0.103 (0.003)\\
 3 & (+4.0, --3.3)  & 600 &  38.1 & --0.064 (0.003) & --0.033 (0.003) & --0.091 (0.003)\\
 4 & (--4.2, +2.2)  & 600 &  37.8 & --0.056 (0.003) & --0.028 (0.003) & --0.091 (0.003)\\
 5 & (+7.2, --11.8) & 1200 &  10.7 & --0.084 (0.004) & --0.029 (0.004) & --0.110 (0.005)\\
 6 & (--11.0, +6.7) & 1200 &   9.5 & --0.082 (0.004) & --0.014 (0.005) & --0.129 (0.005)\\
 7 & (--13.4, --8.3) & 1200 &  16.6 & --0.085 (0.003) & --0.026 (0.003) & --0.149 (0.003)\\
 8 & (--11.0, +11.7)& 600 &   6.6 & --0.092 (0.016) & --0.005 (0.010) & --0.051 (0.013)\\
 9 & (--11.0, +16.7)& 300 &   6.0 & --0.103 (0.014) &  +0.029 (0.014) & --0.078 (0.015)\\
10 & (--11.0, +21.7) & 300 &   6.2 & --0.082 (0.021) & --0.042 (0.014) & --0.039 (0.014)\\
\\

\hline
\\
\multicolumn{8}{l}{$^a$ Offset ($\alpha\cos\delta, \delta$) in $^{\prime}$
relative to $\rm \alpha = 19^h\, 23^m \, 43^s.0$,
$\delta= 14^\circ 30^\prime 38^{\prime\prime}$ (J2000)}\\
\multicolumn{8}{l}{$^b$ On-source integration time}\\
\multicolumn{8}{l}{$^c$ Beam-averaged continuum brightness temperature at 1666 MHz}\\
\multicolumn{8}{l}{$^d$ Line equivalent width ($< 0$ for absorption line), with statistical
errors in parentheses}\\

\end{tabular}}

\end{document}